# The current orbit of Atlas (SXV)

**Demétrio Tadeu Ceccatto[1], Nelson Callegari Jr.[1], Adrián Rodríguez[2]**

[1] São Paulo State University (UNESP), Institute of Geosciences and Exact Sciences, Av 24-A 1515, 13506-900, Rio Claro, Brazil
[2] Observatório do Valongo, Universidade Federal do Rio de Janeiro, Ladeira do Pedro Antônio 43, 20080-090 Rio de Janeiro, Brazil

email: dt.ceccatto@unesp.br

**Abstract.** With the success of the Cassini-Huygens mission, the dynamic complexity surrounding natural satellites of Saturn began to be elucidated. New ephemeris could be calculated with a higher level of precision, which made it possible to study in detail the resonant phenomena and, in particular, the 54:53 near mean-motion resonance between Prometheus and Atlas. For this task, we have mapped in details the domains of the resonance with dense sets of initial conditions and distinct ranges of parameters. Our initial goal was to identify possible regions in the phase space of Atlas for which some critical angles, associated with the 54:53 mean motion have a stable libration. Our investigations revealed that there is no possibility for the current Atlas orbital configuration to have any regular behavior since it is in a chaotic region located at the boundary of the 54:53 mean-motion resonance phase space. This result is in accordance with previous works (Cooper et al. 2015, Renner et al. 2016). In this work, we generalize such investigations by showing detailed aspects of the Atlas-Prometheus 54:53 mean-motion resonance, like the extension of the chaotic layers, the thin domain of the center of the 54:53 resonance, the proximity of other neighborhood resonances, among other secondary conclusions. In particular, we have also shown that even in the deep interior of the resonance, it is difficult to map periodic motion of the resonant pair for very long time spans.

## 1. Introduction

The natural satellites of Saturn display a variety of orbital configurations and unique topological features that have intrigued astronomers, physicists and mathematicians for severalyears (e.g. Peale 1999). Such dynamic environment is responsible for the most diverse gravitational disturbances such that tides, short and long periodic perturbations. In particular, small satellites with their mean radius of the order of ten kilometers or smaller suffer non-negligible quasi-periodic variations in their orbits due to mutual gravitational interactions and, this phenomena imply that many pairs of satellites have their orbits close to commensurability ofthe mean-motions. In this work, we study the dymanics of Atlas. According to Thomas and Helfenstein (2020), Atlas is a small satellite with radius of 14.9 km orbiting near the outeredge of Ring A, see Table 2.

Spitale et al. (2006) combining data previously obtained by the Voyager spacecraft, the Hubble Space Telescope and ground-based telescopes determined precise ephemeris for these small worlds, in particular for Atlas. In their studies, Spitale et al. (2006) suggested that Atlas orbit is perturbed due to mutual gravitational interaction with Prometheus generating a 54:53 resonant mean-motion and this can be associated with the presence of the chaotic motion.

Cooper et al. (2015) investigating the existence of the chaotic motion of Atlas suggested by Spitale et al. (2006) by calculating the Fast Lyapunov Indicator (FLI) method to verify the presence of chaos in the Atlas orbit. They analyzed the critical angles associated with the 54:53 resonant mean-motion for a timespan of 20 years. Their integrations showed that the angle associated with the Corotation Eccentric Resonance ($\varphi_{CER}$) oscillates with temporay oscillation with period about 4.92 years and amplitude around $180°$ followed by circulation intervals. In contrast, the angle associated with Lindblad Eccentric Resonance ($\varphi_{LER}$) has temporary oscillation of about 3 years followed by a long period of circulation (see Figure 2 of Cooper et al. 2015).

In an attempt to determine the origin of this alternation between libration and circulation, Renner et al. (2016) studied the Atlas orbit using the CoraLin model proposed by El Moutamid et al. (2014). Your results show that the region of the space occupied by the CER resonance is superimposed by the chaotic region.

Following the contributions of Cooper et al. (2015) and Renner et al. (2016), we will investigate the Atlas orbit looking for initial conditions that enable stability for the libration of the $\varphi_{CER}$ and $\varphi_{LER}$ angles. In additions, we identiffy possible causes for instability described in Cooper et al. (2015).

**Table 1.** Physical constants for Saturn obtained from Cooper et al. (2015)[a], Horizons[b], $J_2$, $J_4$ and $J_6$ a dimensionless coefficients of Saturn potential expansion.

| Constant | Numerical value |
|---|---|
| $GM^a$ (Km$^3$s$^{-2}$) | 3.7931208x10$^7$ |
| Equatorial radius[b] (Km) | 60268±4 |
| $J_2^a$ | 16290.71x10$^{-6}$ |
| $J_4^a$ | -935.83x10$^{-6}$ |
| $J_6^a$ | 86.14x10$^{-6}$ |

## 2. Methodology

In this work, we follow the methodology given in Callegari and Yokoyama (2010, 2020) and Callegari et al. (2021) and numerically integrate the exact equations of the motion for a system of $N$ satellites mutually disturbed orbiting under the action of the main terms of the Saturn's potential expanded up to second order. In our simulations, for brevity we show the main results considering only $N = 2$, that is, a system formed by Atlas and Prometheus. The justification for this choice stems from Atlas orbital elements can well be determined over the influence of Prometheus, without considering the other effects arising from the perturbation caused by Pandora or other companion satellite (Spitale et al. 2006, Cooper et al. 2015 and Renner et al. 2016). It is worth to note that we have considering more general models in this work, but for brevity only the result within the domains of the three-body problem are shown here.

Two different models were used: i) the system of exact differential equation (Eq. 1-5) present in Callegari and Yokoyama (2010), where the equations of motion are integrated under the influence of the terms $J_2$ and $J_4$ and; ii) direct application of the Mercury package (Chambers 1999) with the addition of the term $J_6$. In both cases i) and ii) we apply the Everhart code "RA15" to solve systems of ordinary differential equations (Everhart 1985).

Physical parameters for Saturn and initial conditions for Atlas and Prometheus provided by the *Horizons* system of ephemerides (http://sdd.jpl.nasa.gov/horizons.cgi) are listed in Tables 1 and 2, respectively. The masses were obtained from Thomas and Helfenstein (2020) and the details for the numerical simulations are described in the caption of the respective figure.

Often, some oscillating elements obtained through numerical integration can be highly influenced by the term $J_2$, causing a fast frequency component in the pericenter ($\omega$), a fact that makes difficult its interpretation (Greenberg, 1981). Thus, as an additional effort, we calculate the geometric elements through the direct application of the algorithm described by Renner and Sicardy (2006).

**Table 2.** Mass and osculating orbital elements at Epoch 2000 January 1 00: 00: 00.0000 UTC computed from Ephemeris system - *Horizons*, in December 13, 2019.

| | Mass (Kg) | a (Km) | e | I (°) | $\omega$ (°) | $\Omega$ (°) | n (°/day) |
|---|---|---|---|---|---|---|---|
| Atlas | 5.75x10$^{15}$ | 138325.32 | 0.00591 | 0.00419 | 200.7 | 235.45 | 592.63 |
| Prometheus | 1.5x10$^{16}$ | 140246.44 | 0.00252 | 0.00801 | 201.,36 | 309.14 | 581.87 |

## 3. The current orbit of Atlas and the 54:53 Prometheus-Atlas mean-motion

Atlas finds itself orbiting a region of dynamic complexity. The understanding of the 54:53 resonant mean-motion make it necessary to determine the stability of its orbit.

## 3.1 The critical angles $\varphi_{CER}$ and $\varphi_{LER}$

Resonant phenomena are common among satellites of Saturn, in particular the Corotation Eccentricity Resonance (CER) and Lindblad Eccentricity Resonance (LER). According to Callegari et al. (2021) the CER resonance occurs when the conjunction between disturber (larger body) and particle (smaller body) happens near the pericenter of the disturbed one, on the other hand, the LER resonance occurs when the conjunction occurs near the pericenter of the particle. The knowledge of the physical behavior of these two angles makes it possible to predict where the conjunctions between the two bodies occur, in addition to verifying the orbital stability of the system.

For the 54:53 resonant mean-motion between Prometheus and Atlas, the angles $\varphi_{CER}$ and $\varphi_{LER}$ can be written as an angular combination of the form $\varphi_{CER} = 54\lambda_P - 53\lambda_A - \varpi_P$ and $\varphi_{LER} = 54\lambda_P - 53\lambda_A - \varpi_A$, where $\lambda = \varpi + l$ represents the mean longitude, $\varpi$ the pericenter longitude, $l$ the mean anomaly, whereas $A$ and $P$ represent Atlas and Prometheus, respectively. Physically, the libration of the $\varphi_{CER}$ angle around 180° means that the conjunctions occur close to the apocenter of Prometheus. In the case of the current orbit of Atlas, Fig. 1 shows that there are periods of alternation between oscillation and circulation of $\varphi_{CER}$.

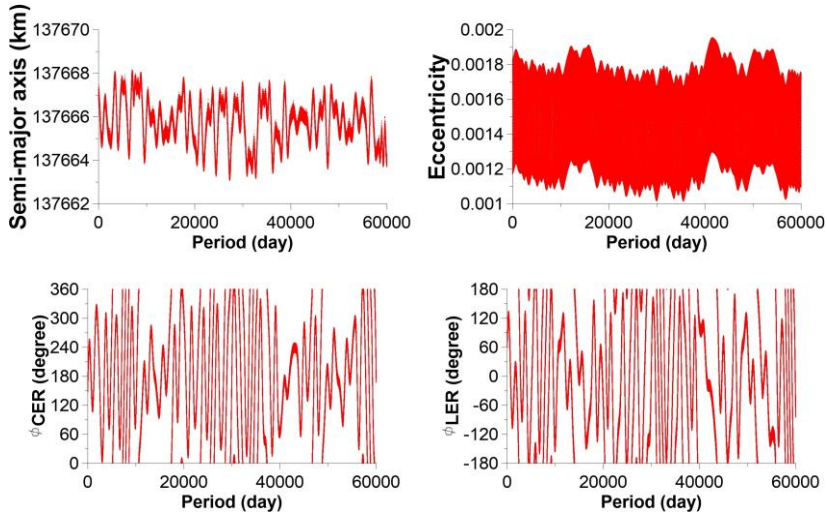

**Figure 1.** Simulation performed with the initial conditions given in Table 2 for an Atlas-like satellite. Semi-major axis and eccentricity geometric elements and critical angles in geometric elements associated with CER and LER type resonances. Note the lack of stability for the oscillation of $\varphi_{CER}$ and $\varphi_{LER}$. The equations were integrated for 60,000 days with a 0.06 day step using the Mercury package.

On the other hand, the $\varphi_{LER}$ angle has a brief period of oscillation, about 5,000 days, followed by a long period of circulation, as shown in Fig. 1.

Cooper et al. (2015) state that Atlas could be added to the list of satellites that have both critical angles librating, however, for a satellite similar to Atlas, we cannot make such an affirmation, as we do not obtain stability for the dynamics of these angles as it occurs for Anthe (Callegari and Yokoyama, 2020) and Methone (Callegari et al. 2021). See also El Moutamid et al. (2014).

Expanding the investigation done by Cooper et al. (2015) we will look for a possible orbital configuration that can guarantee the stability of the libration of $\varphi_{CER}$ and $\varphi_{LER}$. For this task, we will use the phase space mapping through Fourier spectra, a methodology described in Callegari and Yokoyama (2010, 2020) and Callegari et al. (2021).

## 3.2 The mapping of the 54:53 Prometheus-Atlas resonant mean-motion

We will now explore the resonant phase space domain. The mapping is carried out considering the domain of frequencies obtained with the Fourier spectrum for a set of numerically integrated orbits, whose initial conditions are close to the real orbit of Atlas.

Callegari and Yokoyama (2010, 2020) and Callegari et al. (2021) analyze the phase space around the orbital neighborhood of a satellite according to the value of the spectral number N. For the construction of this N it is considered an established value, usually 5% of the reference amplitude for the Fourier spectrum for each of the individual orbits of these test satellites. Then, N provides the number of peaks in the spectrum that are greater than or equal to the pre-set reference value.

We will now investigate the dependence of the resonant motion with the semi-major axis and the eccentricity for an Atlas-like satellite. Fig. 2 shows two dynamic maps built on a dense grid of initial conditions for $(a_0,e_0)$. The red star represents the initial condition $(a_0,e_0)$=(138325.32 km, 0.0059) for Atlas on the initial date (see Table 2).

The map at the left in Fig. 2 represents the phase space of the spectral domain for the semi-major axis and the map on the right represents the spectral domain for the orbital eccentricity. We can identify four distinct regions:

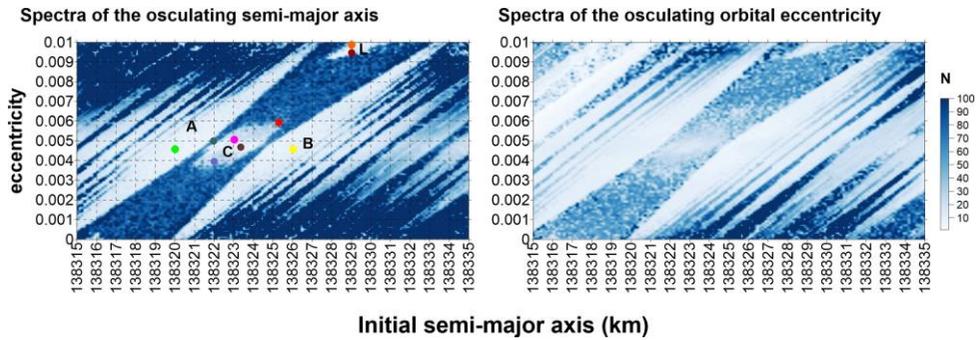

**Figure 2.** Dynamic map constructed from the spectrum of the osculating semi-major axis (left) and orbital eccentricity (right) of 15,000 Atlas-like satellites (numerical scheme i). N is the spectral number with 5% reference amplitude. **C** and **L** are the regions in which $\varphi_{CER}$ and $\varphi_{LER}$ angles have periods of oscillation of approximately 30,000 days and 15,000, respectively, but alternating their behavior. For this simulation, Prometheus and Atlas were used, the set of differential equations had been integrated for 258 years with an interval of 0.18 days. Initial conditions are given in Table 2. The initial element of the real Atlas are indicated by a red star. Different full discs correspond to initial conditions $(a_0,e_0)$ of the orbits shown in Figure 4.

a) A light "blue eye" region, with N less than 30, in the range [138321 km, 138326km]x[0.004, 0.006] indicated by **C**. In this region all angles $\varphi_{CER}$ are oscillating around 180° but the alternation occurs between the oscillation and circulation. However, for the initial conditions (138322 km, 0.004) blue disc and (138323 km, 0.005) magenta disc, in Fig. 2, theangle $\varphi_{CER}$ is librating, see Fig. 3 items (b), (e) and (f). This would probably be the region of the corotation zone associated to the 54:53 Prometheus-Atlas mean-motion resonance, but, the absence of stability in the libration for the other initial conditions, Fig. 3 (f), does not allow this region to be considered as such.

b) The dark blue region, which fills practically the entire map, where the spectrum of the test satellites has a large value for N. In general, such behavior can be associated with chaotic orbits and strong or irregular disturbances in their motions (Callegari and Yokoyama, 2020).

c) Region between [138327 km, 138330 km]x[0.009, 0.01], indicated by **L**. Location where the angle $\varphi_{LER}$ shows oscillation around 0°. The absence of stability in the libration does not allow this region to be considered as such, see Fig. 3 items (d) and (g).

d) Bands A and B: in this region angles $\varphi_{CER}$ and $\varphi_{LER}$ show themselves circulating in a retrograde direction for a brief period of time, about 500 days (band A) and for region B, the angles are circulating in a prograde direction with a period of approximately 400 days. See Fig. 3 items (a) and (c).

We can verify, with the help of the dynamic map given in Figure 2, that there is no region with $(a_0,e_0)$ in which there is stability for the critical angles $\varphi_{CER}$ and $\varphi_{LER}$ libration, neither for both critical angles.

We can observe that for the initial conditions given in Table 2, Atlas is found close to the edge of a possible chaotic region and this is probably responsible for the alternations observed in the angles $\varphi_{CER}$ and $\varphi_{LER}$ observed in Cooper et al. (2015) (See Figure 1).

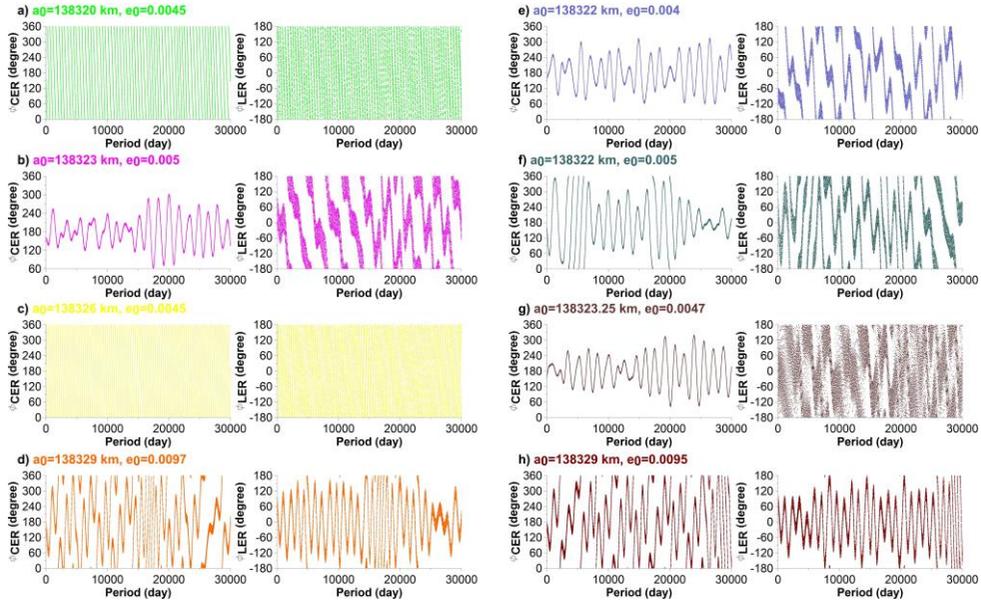

**Figure 3.** Geometrics angle $\varphi_{CER}$ and $\varphi_{LER}$ obtained with the numerical simulations (numerical scheme ii) corresponding to the colored discs in Fig. 2. In a) and c) both are circulating; b) e) and g) the angle $\varphi_{CER}$ are librating and $\varphi_{LER}$ circulating; and g) alternation between libration and circulation of angle $\varphi_{CER}$ and circulation of angle $\varphi_{LER}$; d) and h) libration of angle $\varphi_{LER}$ and circulation of angle $\varphi_{CER}$.

El Moutamid et al. (2014) have applied a general development of corotation and Lindblad resonaces (CoraLin model), ideal to be analyzed with the surface of section technique. The axes of the section are given by a variable $\chi$, representing the resonance width (see Table 1 in El Moutamid et al. (2014)) which is plotted against the corotation angle $\varphi_{CER}$.

In Fig. 4, we show several plots of $\chi$ x $\varphi_{CER}$ for some orbits obtained numerically through the Mercury package taken in the regions (a), (c) and (d), where we consider the full values of $\chi$ obtained from numerical orbits, not the sections. Also we utilize $a' = 137665.519$ km as the reference geometric center of the corotation zone (Renner et al. 2016).

We have the following conclusions:

A) When values for the semi-major axis are between 138322 and 138323 km, $\chi$ assumes small values while $\varphi_{CER}$ alternates between libration and circulation. See Figure 5 in Renner et al. (2016) and Figures 3 and 4 of this work.

B) For large values of eccentricity, $\chi$ becomes diffuse in the $\chi$ x $\varphi_{CER}$ plane, while $\varphi_{CER}$ is circulating.

## 4. Final considerations

The use of the dynamic map obtained through the spectrum of individual orbits revealed that:

i) Atlas is located on the edge of a possible chaotic region. Physically, this location implies the absence of stability for critical angles $\varphi_{CER}$ and $\varphi_{LER}$ found by Cooper et al. (2015);

ii) for the interval of semi-major axis and eccentricity values given by [138322 km, 138323 km] x [0.004, 0.005] the map reveal an initial conditions ($a_0$, $e_0$) in which we can find a stable libration for $\varphi_{CER}$ and also, initial conditions in which the angle $\varphi_{CER}$ presents episodes of libration oscillation followed by circulation;

iii) for eccentricity values greater than 0.009, and semi-major axis between 138328 and 138330 km we identified the region in which periods of libration occur for the angle $\varphi_{LER}$, as well as in ii) we did not find a region our intial condition in which there is stable libration for

$\varphi_{LER}$.

iv) for regions A and B in Fig. 2, the orbits were presented as circulating according to El Moutamid et al. (2014). Such feature indicates that the Atlas-like particle is far from $a'$.

Therefore, we conclude that Atlas is not deeply inserted in to the 54:53 resonance, due to the lack of stability in the $\varphi_{CER}$ and $\varphi_{LER}$ angle libration. This fact implies that Atlas cannot be added to Saturn's list of natural satellites that have both critical angles librating.

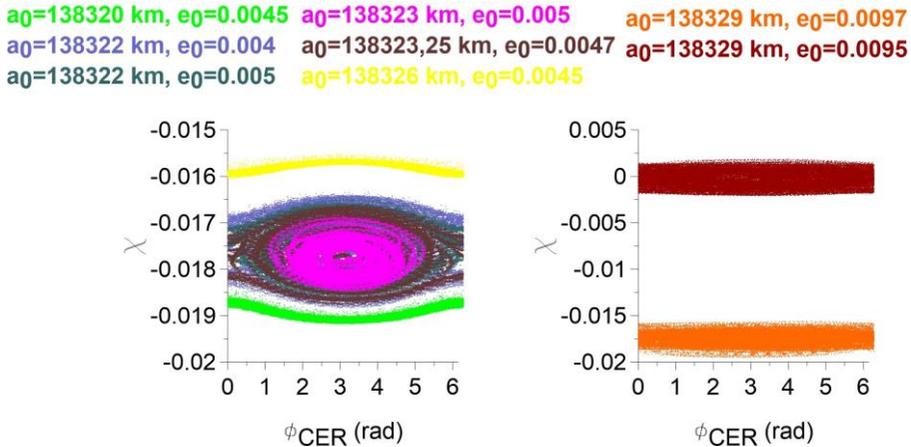

**Figure 4.** Plot for $\chi$ x $\varphi_{CER}$ of several orbits with initial conditions gives in coloured full discs in Fig. 2.